\documentclass[reprint,
superscriptaddress,
amsmath,
amssymb,
aps,
prb,
nobibnotes
]{revtex4-2}

\usepackage{lipsum}
\usepackage{graphicx}
\usepackage{dcolumn}
\usepackage{bm}
\usepackage{comment}
\usepackage{upgreek}
\usepackage[usenames,dvipsnames]{color}
\usepackage[mathlines]{lineno}
\usepackage{multirow}
\usepackage{siunitx}
\usepackage[colorlinks]{hyperref}
\usepackage{amssymb,amsmath,amsthm,enumitem}
\usepackage{soul}
\usepackage{nameref}
\usepackage{chngcntr}

\DeclareUnicodeCharacter{2212}{-}
\DeclareGraphicsExtensions{.pdf,.png,.jpg}

\newcommand{\hpar}{{H$_\parallel$}}
\newcommand{\hperp}{{H$_\perp$}}

\definecolor{bluegray}{RGB}{40,180,160}
\definecolor{navygray}{RGB}{110,140,170}
\definecolor{meadowgreen}{RGB}{0,128,0}
\definecolor{magenta}{RGB}{255,0,255}
\definecolor{lightgrey}{RGB}{200,200,200}
\definecolor{C0}{RGB}{31,119,180}
\definecolor{C1}{RGB}{255,127,14}
\definecolor{C2}{RGB}{44,160,44}

\hypersetup{
citecolor={bluegray}, 
linkcolor={navygray}, 
}

\begin{document}

\title{Hydrogen crystals reduce dissipation in superconducting resonators}

\author{Francesco~Valenti}
\thanks{First two authors contributed equally.}
\affiliation{IQMT,~Karlsruhe~Institute~of~Technology,~76344~Eggenstein-Leopoldshafen,~Germany}
\affiliation{Current address: IBM Quantum, IBM T. J. Watson Research Center, Yorktown Heights, 10598 NY, USA}

\author{Andrew~N.~Kanagin}
\thanks{First two authors contributed equally.}
\affiliation{Vienna~Center~for~Quantum~Science~and~Technology,~Atominstitut,~TU~Wien,~1020~Vienna,~Austria}

\author{Andreas~Angerer}
\affiliation{Vienna~Center~for~Quantum~Science~and~Technology,~Atominstitut,~TU~Wien,~1020~Vienna,~Austria}

\author{Luiza~Buimaga-Iarinca}
\affiliation{INCDTIM, 400293~Cluj-Napoca, Romania}

\author{\\Cristian~Morari}
\affiliation{INCDTIM, 400293~Cluj-Napoca, Romania}

\author{Jörg~Schmiedmayer}
\affiliation{Vienna~Center~for~Quantum~Science~and~Technology,~Atominstitut,~TU~Wien,~1020~Vienna,~Austria}

\author{Ioan~M.~Pop}
\email{ioan.pop@kit.edu}
\affiliation{IQMT,~Karlsruhe~Institute~of~Technology,~76344~Eggenstein-Leopoldshafen,~Germany}
\affiliation{PHI,~Karlsruhe~Institute~of~Technology,~76131~Karlsruhe,~Germany}
\affiliation{Physics~Institute~1,~Stuttgart~University,~70569~Stuttgart,~Germany}

\begin{abstract}
We show that the internal quality factors of high impedance superconducting resonators made of granular aluminum can be improved by coating them with micrometric films of solid para-hydrogen molecular crystals. We attribute the average measured $\approx8 \%$ reduction in dissipation to absorption of stray terahertz radiation at the crystal-resonator interface and the subsequent dissipation of its energy in the form of phonons below the pair-breaking gap. Our results prove that, contrary to expectations, replacing the vacuum dielectric atop a superconducting resonator can be beneficial, thanks to the added protection against Cooper pair-braking terahertz radiation. Moreover, at the level of internal quality factors in the $10^5$ range, the hydrogen crystal does not introduce additional losses, which is promising for embedding impurities to couple to superconducting thin-film devices in hybrid quantum architectures.
\end{abstract}

\maketitle

\section{Introduction}

%intro : SC and QP
 Superconducting circuits play a central role in detection \cite{day2003broadband}, metrology \cite{taylor1989new, Crescini2023}, and quantum hardware \cite{arute2019quantum, Krinner2022, Sivak2023}, wherein superconducting quantum bits (or qubits) are a scalable technology that leverages Josephson junctions as low-loss sources of nonlinearity. While their development during the last decades has been impressive, decoherence is still the Achilles' heel of superconducting qubits, e.g. hindering efforts towards quantum error correction \cite{fowler2012towards}. One of the sources of decoherence are broken Coopers pairs, i.e. quasiparticles (QPs) \cite{aumentado2004nonequilibrium, barends2008quasiparticle, shaw2008kinetics, catelani2011quasiparticle, de2011number, riste2013millisecond, wang2014measurement, nsanzineza2014trapping, serniak2018hot}. Operating circuits in dilution cryostats, in the mK range, ensures the absence of QPs of thermal origin. Sources of the remaining, athermal QPs include high energy particles \cite{Vepslinen2020impact, Wilen2021, McEwen2022} and stray terahertz radiation; in turn, mitigation strategies include methods to abate both incoming high energy particles \cite{valenti_cardani_demetra} and the athermal phonons subsequently generated in the substrate \cite{HenriquesValenti2019}, as well as terahertz shielding \cite{barends2011minimizing, Corcoles2011, serniak2018hot, Connolly2023}. 

% hybrid architectures
Compared to macroscopic superconducting circuits, microscopic quantum objects such as electrons and atoms couple less strongly to their environment, leading to increased coherence times, but also making up-scaling of their control challenging. An enticing prospect is the realization of hybrid architectures that would exploit the advantages of each implementation | microscopic quantum memories controlled by macroscopic quantum gates. To that end, various solutions have been proposed and implemented such as coupling superconducting circuits to atoms \cite{hybrid_atoms,Verdu2009a}, molecules \cite{hydrid_mols,Andre2006}, and electrons \cite{hybrid_spin, Albertinale2021, Wang2023}, and a whole range of cavity QED experiments were realized for dense ensembles of magnetically coupled vacancies \cite{hybrid_nv,Amsuss2011,Putz2014,Putz2016,Angerer2017,Astner2018}. 

%Vienna's choice of hybrid arch
The coupling itself can be either magnetic or electric. Working with superconductors and magnetic systems is fundamentally hindered by the presence of a critical magnetic field for superconductors, and possibly removes magnetic tuning as a reliable control knob for the superconducting circuit parameters. Low loss, high impedance superconducting resonators bear the promise of being able to couple electrically to microscopic species instead, creating more robust and versatile hybrid systems \cite{Samkharadze2015high, scarlino2019coherent, scarlino2022situ}. In particular, a promising hybrid architecture is a high impedance superconducting resonator coated with a noble gas crystal: such a soft, low-stress, spin-zero matrix would be a versatile host for embedding dipole-bearing impurities \cite{upadhyay2016longitudinal,kanagin2018design, upadhyay2020ultralong}, with strong magnetic coupling recently demonstrated between Na spins in a Ne matrix and a superconducting resonator \cite{kanagin2023priv}.

% this work
In this work, we present results that are relevant to both the developmental avenues of improving existing Josephson junction-based qubits and towards the realization of hybrid quantum systems based on superconducting resonators with noble gas matrices as an impurity host. We show that we can control the growth of a spinless, molecular crystal of para-hydrogen atop high impedance granular aluminum resonators, thereby providing a magnetically inert matrix to embed microscopic species to act as quantum memories. Furthermore, we show that such crystals do not significantly increase dielectric loss in the measured resonators, and \textit{decrease} quasiparticle losses when the film is exposed to pair-breaking terahertz radiation, hinting to a nontrivial surface effect. 

\begin{figure*}[t!]
  \def\svgwidth{1\textwidth} 
    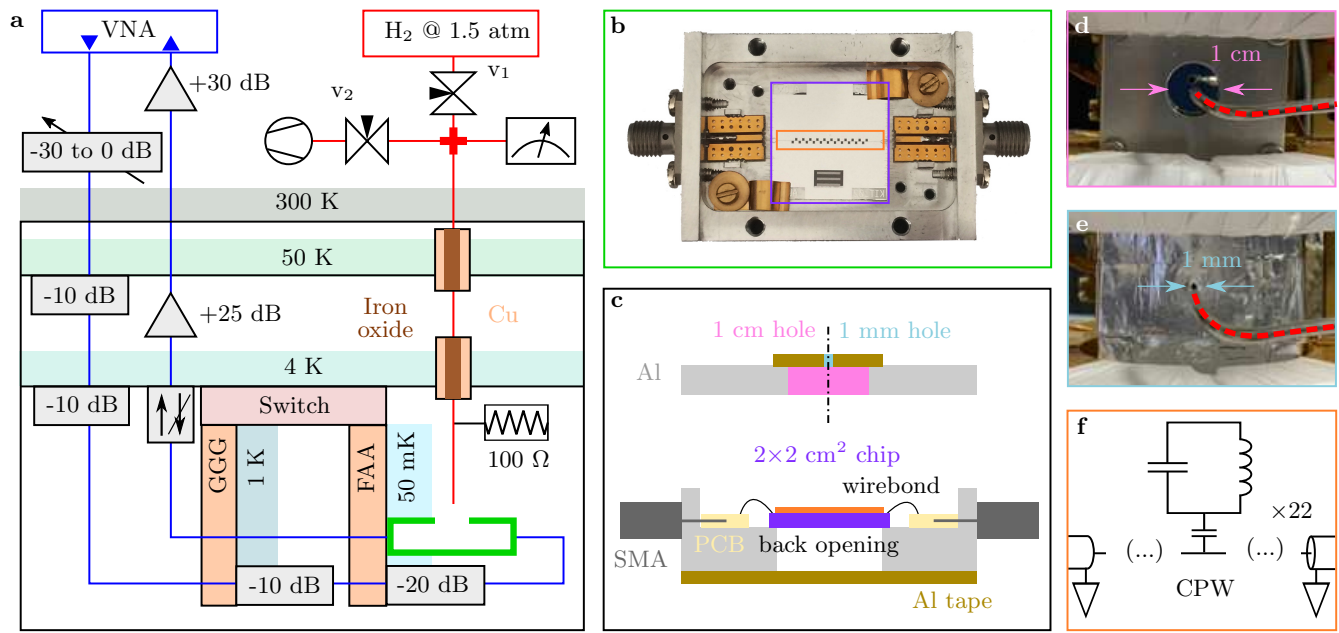
    \caption{\textbf{Cryogenic, microwave, and gas handling setup. a}, schematics of the adiabatic demagnetization refrigerator together with microwave (blue) and gas (red) lines. A pulse tube cryocooler and liquid helium are used to cool down successive thermal stages at 50 K and 4 K, respectively. The latter is connected via a thermal switch to two separate copper rods filled with two different paramagnetic salts: Gallium Gadolinium Garnet (GGG) and Ferric Ammonium Alum (FAA), with magnetic ordering temperatures of about $1$ K and $50$~mK, respectively. The sample holder is mounted on the FAA-filled rod. A vector network analyzer (VNA) is used to perform transmission microwave spectroscopy on the sample. The signal towards the colder stages is progressively attenuated in order to reduce its thermal noise; after transmission through the sample, it is routed by an isolator and amplified both at cryogenic and room temperature stages. The first needle valve (v$_1$) is used to regulate the incoming gas pressure from a room temperature tank. The line is connected via a four-way joint to a digital barometer and to a pump, regulated with an additional needle valve (v$_2$), used to extract gaseous residues between successive crystal growths. The gas entering the cryostat is fed through copper tubes thermalized at 50~K and 4~K and filled with an iron oxide for conversion to the para spin isomer of H$_2$ (see main text for discussion). During deposition the second tube is heated from 4 to 14~K by feeding a PID-controlled current to a 100~$\Omega$ load in order to promote gas flow. \textbf{b}, photograph of the solid aluminum transmission sample holder showing the two SMA ports and printed circuit boards used to implement impedance-matched connections to the on-chip feedline (cf. Fig. F1 in Ref.~\cite{valenti2019interplay} for details). \textbf{c}, schematic half-section of the sample holder and its lid. Note the $1.5\times 1.5 \;\mathrm{cm}^2$ opening on the backside of the sapphire chip (sealed with aluminum tape) and the circular hole in the lid, $1$~cm in diameter. This is either left as is (panel \textbf{d}) or covered with two layers of aluminum tape with a $1$~mm hole poked through (panel \textbf{e}). The stainless steel gas-carrying tube visible in both panels \textbf{d} and \textbf{e} is highlighted in red (cf.~panel \textbf{a}). \textbf{f}, schematics of the circuit, consisting of a coplanar waveguide (CPW) feedline capacitively loading 22 resonators (cf.~Fig. \ref{fig_loading} for details on feedline and resonator geometry).} \label{fig_setup}
\end{figure*}

%this manuscript
This manuscript is organized as follows: in Section \ref{sec_setup}, we report on the cryogenic setup used in this experiment, with the ability of growing noble gas crystals atop the chip; in Section \ref{sec_samples}, we describe the samples and their parameters; in Section \ref{sec_crystals}, we discuss the deposition and characterization of the para-hydrogen crystals; and, in Section \ref{sec_model}, we present simulation results to interpret their beneficial effects. In-depth discussions can be found in the \hyperref[sec_suppl]{Appendices}.

\begin{figure*}[t!]
  \def\svgwidth{1\textwidth} 
    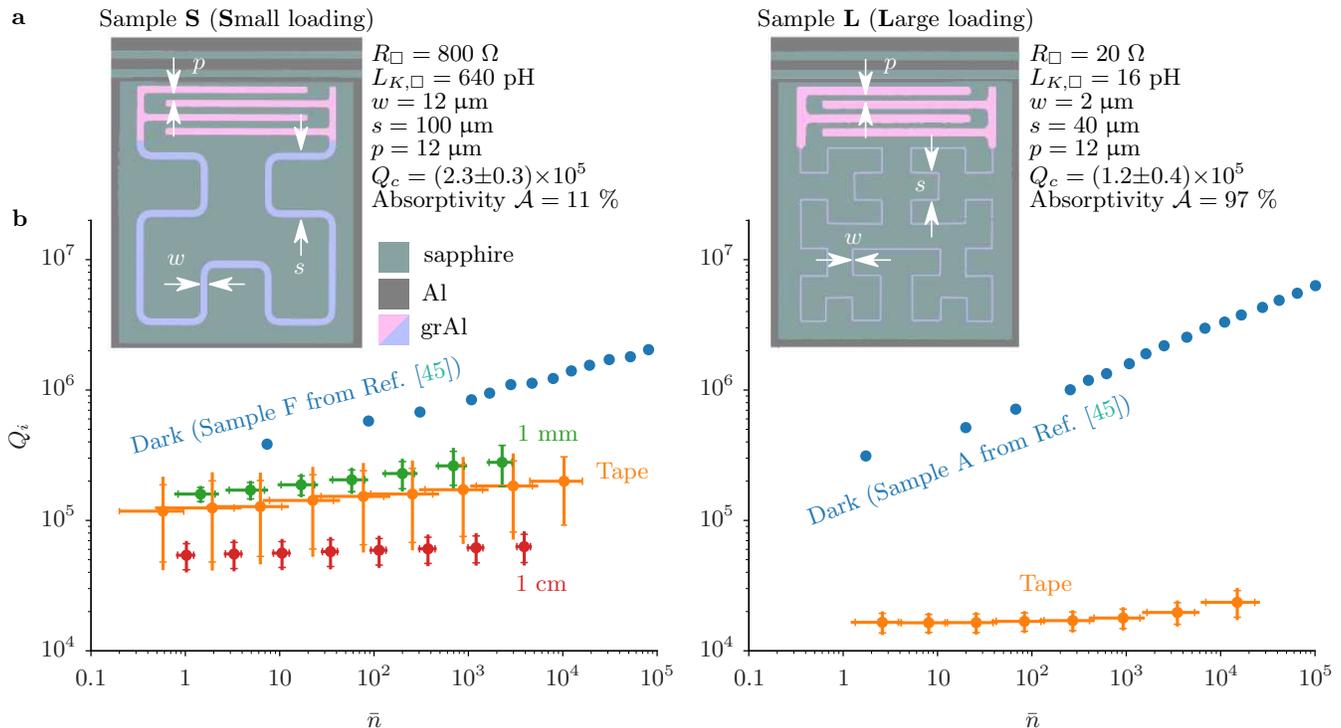
    \caption{ \textbf{Terahertz loading and quasiparticles. a}, false-colored micrographs of the granular aluminum (grAl) resonator types used in the two measured samples, dubbed ``S'' and ``L'' for Small and Large coupling to vacuum impedance, respectively. The grAl resonators (pink/violet), patterned on sapphire (sage) and loaded by a CPW feedline with aluminum pin and ground plane (gray), share the same interdigitated capacitor design (pink), with variable finger length covering the $40-60$~fF range. The grAl inductors (violet) are Hilbert curves of degree 2 and 3 respectively (see main text for discussion), having different widths to guarantee a comparable total inductance despite the factor 40 difference in resistivity. The resulting coupling to vacuum (cf.~Eq.~\eqref{eq_matching}) gives absorptivities $\mathcal{A}=11\%$ and $97\%$ for sample S and L, respectively. \textbf{b}, internal quality factor $Q_i$ as a function of the average number of drive photons in the resonator. The plotted traces are averaged over all visible resonators | error bars along both axes cover one standard deviation. Progressively increasing the optical loading on the sample, by increasing both the aperture and the coupling to vacuum, decreases the internal quality factor due to terahertz radiation breaking Cooper pairs: the largest $Q_i$ is attained when the sample holder cap (cf.~Fig.~\ref{fig_setup}c) is solid aluminum instead of having a hole and the sample is placed into a terahertz shield, which we dub ``dark'' measurement, as reported in Ref.~\cite{valenti2019interplay}. The $Q_i$ measurements taken with the front-facing hole covered with two layers of aluminum tape (orange) are denoted as ``Tape'', and the $Q_i$ measurements based on the diameter of the front-facing hole, 1~mm (green) and 1~cm (red) | cf. Fig.~\ref{fig_setup}d and e | are plotted as well. Note that, due to the relatively high coupling quality factor, systematic uncertainties introduced by Fano interference~\cite{RiegerGunzler2022} are below the fitting uncertainty | whereas such uncertainties are likely playing a significant role for the data from Ref.~\cite{valenti2019interplay} at higher powers.
    } \label{fig_loading}
\end{figure*}

\section{Setup} \label{sec_setup}

% ADR
In order to grow hydrogen crystals atop superconducting films and characterize them, we devised an adiabatic demagnetization refrigerator (ADR) system, schematized in Fig.~\ref{fig_setup}a, that is modified and optimized in the spirit of Ref.~\cite{kanagin2018design}. ADR technology is chosen over commercial dilution-type cryogenic solutions for its ability to rapidly cycle between warmup (50 K) and cooldown (50 mK), on the order of an hour each way, as well as allowing for growths at various substrate deposition temperatures.

% gas handling
A room temperature reservoir of gaseous molecular hydrogen is connected to the gas piping through a needle valve (v$_1$). The overpressure in the reservoir is sufficient to result in gas flow without the need of an active component. A digital barometer is used to regulate the opening of v$_1$, commonly resulting in $\sim 10$~mbar on the cryostat input. The molecular hydrogen gas at room temperature is a mixture of the three ortho and one para spin isomers, essentially thermally equipartitioned. However, at the low temperature of operation (tens of mK), the expected distribution favors the lower energy singlet state. In order to isolate the singlet state, which is spinless and symmetric, we pass the gas through copper tubes filled with iron oxide (Fe$_2$O$_3$). The porous rust offers a vast effective surface ($\sim 50 \;\text{m}^2$), acting as a catalyst for isomer relaxation towards the singlet state. The gas is cooled from 300~K to roughly 50~K through the first copper tubing stage, and down to 14~K in the second copper tube. At a temperature of 14~K we expect to have a para-hydrogen purity above 99$\%$ \cite{kanagin2018design}.

% heater
A custom built PID controller feeds current to a $100$~$\Omega$ load connected to the last section of the gas tube in order to promote gas flow and maintain temperature stability. The resulting tube temperature reaches $14$ K, while the chip goes from $\sim 50$~mK to $\sim 3$~K. Crystal growth is stopped by closing valve v$_1$. After that, valve v$_2$ is opened and a rotary pump is used to extract gaseous residue in the tube; the heater is still on during this process, and is turned off when v$_2$ is closed and the pump is turned off. 

% sample holder body
The sample holder is depicted and schematized in Fig.~\ref{fig_setup}b and c, respectively. The solid aluminum casing hosts printed circuit boards (PCB) onto which standard SMA microwave connectors are soldered, enabling impedance matched bonding from casing to chip. The design is inherited from the NIKA project~\cite{nika2}: the sample holders sport a $1.5 \times 1.5$~cm$^2$ opening back-illuminating the chip, which enables cooling down in an ``optical'' cryostat to gather performance data under millimeter-wave illumination, as described in e.g. Ref.~\cite{valenti2019interplay}. In this work, the back opening is closed with aluminum tape. On the other hand, we employ aluminum lids with a $1$~cm aperture (cf.~Fig.~\ref{fig_setup}d), in order to allow for crystal growth atop the patterned film. We also measure the same setup with the front hole covered with aluminum tape that had a centered hole poked through (cf.~Fig.~\ref{fig_setup}e), resulting in a $1$~mm aperture. It is important to remember that this allows the gas tube, thermalized to $3-4$ K, to shine black body radiation onto the chip: all measurements in this work pertaining to the effect of crystal growth were taken under a terahertz load that is significantly higher to the one found in a standard dilution cryostat.

\section{Samples}  \label{sec_samples}

% grAl resonators 
Samples in this work are $2\times2\;\text{cm}^2$ sapphire chips hosting 22 resonators each (cf.~Fig.~\ref{fig_setup}f). The resonators are fabricated using granular aluminum (grAl), a composite material made of aluminum nanocrystals in an amporphous aluminum oxide matrix \cite{abeles1966gral,deutscher1973granular}: its resistivity, and kinetic inductance with it, can be tuned from aluminum-like to up to three orders of magnitude larger by controlling the oxygen flow during aluminum deposition \cite{levy_bertrand_collective}. Kinetic inductance fractions approaching unity can be obtained this way, thereby maximizing the participation ratio of quasiparticle losses while maintaining quality factors of order $10^5$ at the single photon regime, and in excess of $10^6$ when strongly driven \cite{Grunhaupt2018QPgral}. The two different films presented in this work, samples S and L (cf.~Fig.~\ref{fig_loading}a), have normal state sheet resistance per square $R_\square = 800\;\Omega$ and $20\;\Omega$, respectively.

% device description
The resonators are lumped element devices consisting of a four-finger interdigitated capacitor and a Hilbert shaped \cite{hilbert1891ueber} inductive meander. The finger length is swept to vary the capacitance from $40$ to $60$~fF. The Hilbert fractal shape is chosen for its property of being space filling while offering an approximately equal distribution of vertical and horizontal segments, rendering the apparatus sensitive to two light polarizations at once \cite{rosch2014development}. The $20$~nm thick grAl is patterned onto a $330$~$\upmu$m thick $c$-plane sapphire wafer with e-beam lift-off lithography, followed by an optical lift-off lithography step to deposit the $50$-nm thick aluminum feedline and ground plane, which are wirebonded with aluminum threads to the sample holder PCBs (cf.~Fig.~\ref{fig_setup}b and c).

% terahertz loading
We are interested in evaluating how much the resonator will absorb the aforementioned black body radiation coming from the gas tube. Following Ref. \cite{valenti2019interplay}, the resonator absorptivity, i.e. the ability of electromagnetic radiation to couple into the film due to the impedance matching between the device and the vacuum, defines a normal state sheet resistance per square that is optimally coupling as 
\begin{equation}
    R_{\square,\text{match}} \approx \frac{w}{s} {Z_0},
\end{equation}
where $w$ is the meander width, $s$ is the size of the zeroth order fractal structure, and $Z_0 \approx 377$~$\Omega$ is the vacuum impedance. The absorptivity corresponding to an impedance mismatch is defined as 
\begin{equation} \label{eq_matching}
    \mathcal{A} = 1 - \frac{|R_{\square,\text{match}}-R_\square|}{R_{\square,\text{match}}+R_\square}.
\end{equation}
We therefore possess two independent tuning knobs to control the coupling of our superconducting films to stray pair-breaking radiation incident on the sample holder: the absorptivity $\mathcal{A}$, given by the interplay of film resistivity and resonator geometry, and the lid aperture (a circle of either $1$ cm or $1$ mm in diameter). We dub the two fabricated samples in this work ``S'' and ``L'' for small ($\mathcal{A} = 11\%$) and large ($\mathcal{A} = 97\%$) absorptivity, respectively.

% analysis of samples with no crystal
We characterize the samples by performing transmission microwave spectroscopy using a commercial Vector Network Analyzer (VNA). The scattering data is fitted using the Qkit \cite{qkitcirclefit} circle fitting routine, based on Ref.~\cite{probst2015efficient} | cf.~Appendix~\ref{sec_circle_fit} for an example of a fit. This allows us to estimate the internal and coupling quality factors of the resonators, as well as the average number of circulating photons at resonance for a given drive power (cf.~Appendix~\ref{sec_phot_num}). We plot the internal quality factor $Q_i$ as a function of the average photon number $\bar{n}$ in Fig.~\ref{fig_loading}b. The internal quality factor increases with the resonator drive. This can be attributed both to saturation of dielectric loss \cite{hunklinger1972saturation,golding1973nonlinear} or activation of quasiparticle diffusion \cite{Grunhaupt2018QPgral}. Increasing the terahertz loading, by both increasing the absorptivity (i.e. from sample S to sample L) and the aperture size (colored markers) has a twofold effect. The average $Q_i$ is decreased, suggesting an increased quasiparticle background. Furthermore, the improvement of $Q_i$ with increasing drive power (i.e., the positive slope of $Q_i$ vs. $\bar{n}$), is decreased as well, further hinting at a constant background of QPs becoming the dominant loss mechanisms.

\begin{figure}[t!]
  \def\svgwidth{1\columnwidth}     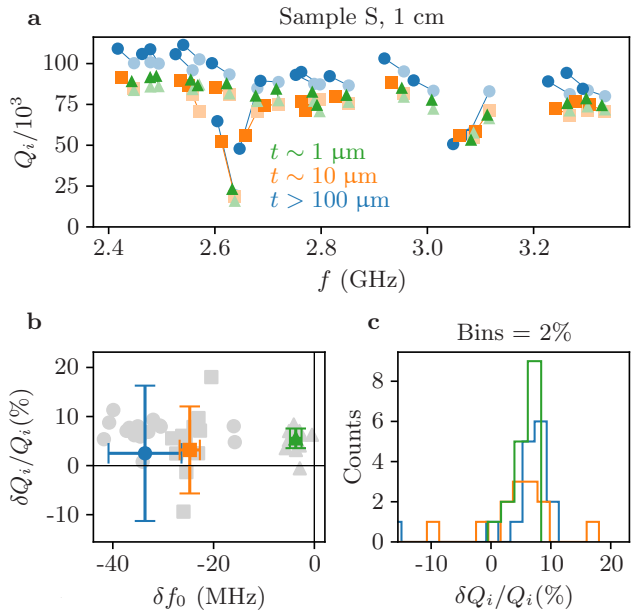
  \caption{\textbf{Effect of crystal thickness on dissipation.} \textbf{a}, internal quality factors $Q_i$ for all visible resonances in sample S, measured with the 1 cm aperture over three separate cooldowns, in which crystals with different thickness $t$ were grown. The reported values are averaged over 10 successive measurements each taken 5 minutes apart; error bars are comparable to marker size and thus omitted. Shaded and full colors represent measurements before and after crystal deposition, respectively. The full dataset is plotted. \textbf{b}, relative $Q_i$ variation as a function of the frequency shift induced by the dielectric change. For clarity, the value (colored markers) is averaged over all values (full gray markers) by dropping the largest and smallest (empty gray markers, when visible). The $y$-axis is bounded by two standard deviations calculated on all values for the deposited three crystals. \textbf{c}, distribution of the relative $Q_i$ shifts, with $x$-axis bounded by two standard deviations calculated on all values. }\label{fig_qi}
 \end{figure}

\section{Crystals}  \label{sec_crystals}

 % crystal deposition
We turn our attention to the effect of adding hydrogen crystals. By varying both the room temperature overpressure of the gaseous hydrogen reservoir and the deposition time, we deposit films with varying thickness, corresponding to circa $1$, $10$, and above $100$~$\upmu$m (cf.~Appendix~\ref{sec_thickness}). In Fig.~\ref{fig_qi}a, we report the internal quality factor measured at $\bar{n} \sim 10^4 \lessapprox n_c$ photons, where $n_c$ is the critical number of photons at bifurcation \cite{eichler2014controlling,valenti2019interplay}, for sample S with the 1 cm aperture. This is the only sample/aperture combination for which it is possible to track the effect of adding crystals: the smaller $1$ mm aperture does not allow for sufficient gas flow and resulting crystal growth (no detectable shift in resonator frequency), while for the case of sample L with the $1$ cm aperture the QP background is too high (resonances are not visible due to increased loss). We measure at the highest available power in order to saturate dielectric loss and render QP loss the dominant dissipation channel.

% baseline and statistics
The reported data were measured over three different cooldowns. In each cooldown, one crystal was grown, and the sample was characterized before and after growth. Several features of interest can be observed. The baseline $Q_i$ value (i.e., before deposition) is constant except for a small downward slope towards higher frequency. The reduced $Q_i$ in the two outliers may be due to either fabrication defects, the dielectric shift bringing the resonator frequency in the vicinity of a more lossy microwave environment, or the resonator being located below the aperture thereby increasing its radiative decay. We wish to point out that we are not equipped to verify the latter hypothesis: not all 22 fabricated resonators are visible in each sample, possibly due to fabrication defects. This, together with the limited precision in reproducing the effective resonance frequency with FEM (cf.~Appendix~\ref{sec_sonnet_simu}), possibly due to oxidation fluctuations in grAl, means we can not robustly map measured resonances to a spatial position.

 \begin{figure}[t!]
  \def\svgwidth{1\columnwidth} 
    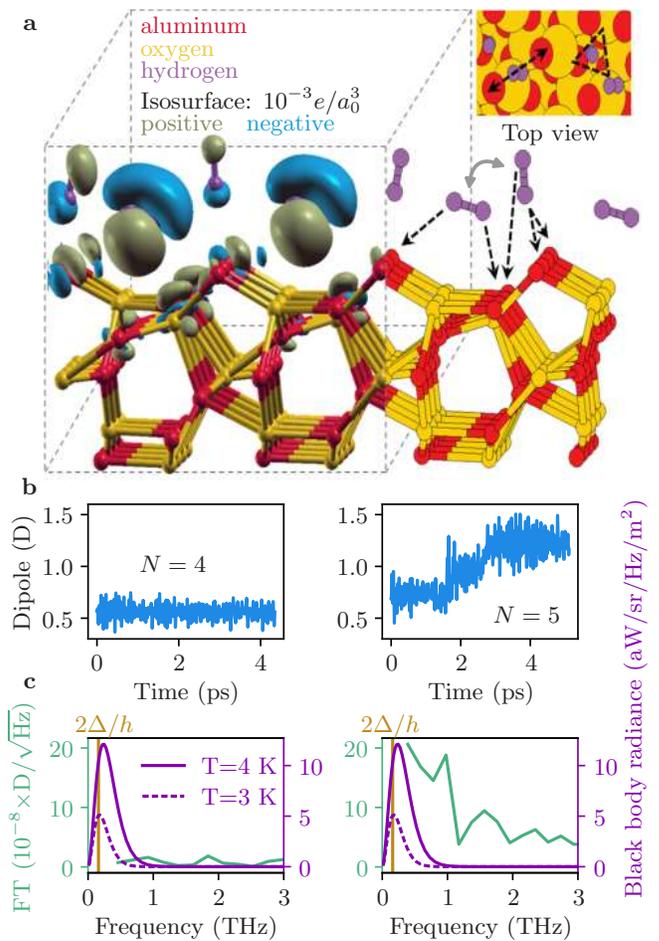
    \caption{\textbf{Terahertz absorption model}. \textbf{a}, snapshot of the supercell used in the density functional theory (DFT) simulations (see main text for discussion), made of aluminum (red) and oxygen (yellow). Eight initially charge-neutral hydrogen molecules (purple) adhere via a site-dependent physisorption mechanism resulting in either charging or discharging (cf.~isosurface representing charge density) and horizontal or vertical configurations, respectively. The H$_2$ molecules in each supercell settle into one of two configurations: \textit{almost perpendicular} to the surface at a $\approx80$\textdegree ~angle, and \textit{almost parallel} to the surface at a $\approx15$\textdegree ~angle. The grey double sided arrow indicates that hydrogen molecules can switch between the parallel and perpendicular configurations, resulting in a change of dipole moment. Black dashed arrows indicate the H-O coordination partners (the bottom H for the perpendicular configuration has two oxygen coordination partners). \textbf{b}, time traces of the total dipole of the supercell with both $N=4$ and $N=5$ adsorbed molecules (cf.~Appendix~\ref{sec_dft}).  \textbf{c}, spectral content of the above time traces (green) together with the blackbody radiance of the gas deposition tube, estimated to be between 3 and 4 K (purple, second $y$-axis). The latter constitutes the dominant electromagnetic background seen by the samples, and is maximum close to the superconducting spectral gap at $160$~GHz (gold). The hydrogen molecule dynamics result in an increased mode density around this region, leading to absorption.} \label{fig_ir}
\end{figure}

% baseline fluctuations
The baseline value also visibly fluctuates across different cooldowns. This is also visible by comparing the values to those of Fig.~\ref{fig_loading}b, left column, at $\bar{n}=10^4$. In order to make sure that it does not fluctuate within a single cooldown, the reported values before and after deposition are averaged over 10 measurements spaced 5 minutes apart: the values proved to be stable, and error bars are not shown since they are too small compared to marker size. A longer observation window was not possible because of the duration of ADR cooldowns, limited to hours. The difference across cooldowns (both in baseline and resonant frequencies) can be explained by a combination of film degradation and small differences in the setup, particularly the position of the gas tube with respect to the sample, resulting in different loading from the tube black body radiation. 

% dielectric shift
The addition of crystals results in a lowering of the resonant frequency. This is expected as we are substituting vacuum with a different dielectric, and this effect, together with FEM simulation, allows us to obtain an estimate for the crystal thickness. Furthermore, the internal quality factor $Q_i$ is increased as a result, which is surprising | at least from a pure material science standpoint, substituting vacuum with any given dielectric should result, if anything, in increased losses. As illustrated in Fig.~\ref{fig_qi}b, the improvement of $Q_i$ seems not to be correlated with the frequency shift, or, in other words, the improvement of $Q_i$ is not correlated with the thickness of the deposited crystal. This suggests the cause of the increase in $Q_i$ to be purely a surface effect. Furthermore, by examining Fig.~\ref{fig_qi}c, the data suggest that best results come from the thinnest crystals: the resulting improvement of $Q_i$ is centered around $\lessapprox 8\%$ | comparable to that of thicker crystals, but with a sharper distribution. This hints at thicker crystals starting to introduce more expected losses associated with potential defects in their structure. Note that we show a comparison of $Q_i$ at high power (Fig.~\ref{fig_qi}a) rather than a full trace of $Q_i$ v. $\bar{n}$ (as e.g. in Fig.~\ref{fig_loading}b) because of time constraint reasons: a full trace takes longer to measure, especially at low photon number when longer averaging is required. This was incompatible with the ADR cooldown time and the necessity of performing several successive measurements to check the time stability of $Q_i$. In future work, the measurement of the full trace could reveal further information about the microscopic mechanisms involved. 

\section{Model} \label{sec_model}
 
% DFT
In order to elucidate a mechanism for the quality factor improvement, we perform DFT simulations on a simplified model of our system (cf.~Appendix~\ref{sec_dft}). Since the experimental results indicate a pure surface effect as the likely cause of the observed performance improvement, we disregard the granular nature of our films and focus on microscopic phenomena at the interface between the para-hydrogen crystal and the resonators. A snapshot of the alumina supercell under study is reported in Fig.~\ref{fig_ir}a, together with eight H$_2$ molecules. The incoming molecular hydrogen has been purified into its para spin isomer, which is a charge neutral species thanks to its symmetric wavefunction. By starting a dynamic simulation with the hydrogen molecules away from the cell, we can observe how they adhere to the oxide in two distinct ways: either perpendicular, resulting in the molecule charging, or parallel to the surface, resulting in the molecule discharging. As such, the surface becomes enriched with loosely bound dipoles. Since the model is microscopic at the level of a few atoms, we expect that the granularity of the grAl surface, on length scales of nanometers, does not significantly change the above picture. 

% charge modes
The total dipole of the supercell evolves over time, in a fashion that is reminiscent of telegraphic noise | albeit switching through several possible values, each corresponding to a certain hydrogen molecule switching configuration: we report a section of the time-evolving dipole in Fig.~\ref{fig_ir}b. This is the key element of the beneficial effect of hydrogen crystals against electromagnetic radiation: adding hydrogen crystals results in increasing the density of charge modes in the supercell. More specifically, such density of modes (computed as the Fourier component of the simulated time evolution of the supercell dipole) is maximal in the low THz region of the spectrum, as reported in Fig.~\ref{fig_ir}c. This is comparable with two other relevant quantities: the peak of the blackbody radiation coming from the gas tube, thermalized in the $3-4$ K range, and the spectral gap of the resonators.

% energy dumping
Less than 1\% of the incoming radiation is reflected at the hydrogen-vacuum interface (cf.~Appenidx \ref{sec_thickness}). Electromagnetic energy in the low THz range is coupled into the flip-flop motion of the hydrogen molecules adsorbed onto the surface. This energy can be dissipated in three main ways, as either rotation, translation or vibrations of the molecules | the latter being able to couple to phonons in the underlying film. We posit that the the energy that is transferred vibrationally to the lower layers has to overcome an inefficient mechanical coupling between the hydrogen crystal (a soft Van der Waals structure), the outer alumina (a hard amorphous oxide), and the actual granular film of the device. Thus, part of the energy absorbed at the interface between the oxide and the hydrogen crystal cannot reach the superconductor and it is scattered below the pair-breaking threshold, resulting in decreased QP losses. We wish to note the speculative nature of this argument: follow up work on this topic should focus on a understanding of the mechanical coupling between layers, giving a tool for a quantitative budgeting of the phonon mediated energy transfer. In addition, further work should explore other potential mechanisms of performance improvement, such as e.g. the deposited crystal inducing a favorable reconfiguration of the landscape of possible two-level system fluctuators on the surface \cite{Muller_2019}.
\section{Conclusions}

% results summary 
We have shown that high impedance grAl resonators can be coated with a micrometric layer of magnetically inert para-hydrogen molecular crystal, suitable for embedding microscopic species. Interestingly, the crystal does not add measurable dielectric loss, and results in decreased quasiparticle loss when the film is exposed to a $3-4$ K blackbody. In particular, the internal quality factor of the resonators is consistently increased by around $8\%$ indiscriminately for crystals in the $1-100$~$\upmu$m thickness range.

% analysis summary
We show DFT simulations of a supercell consisting of aluminum oxide atop of pure aluminum with para-hydrogen molecules approaching the surface. The molecules show charging or discharging depending upon physisorption geometry. Dipole fluctuations coming from instability of the adsorbed configuration over time result in an increased charge mode density around the peak of the blackbody emission, which is in turn close to the spectral gap of the film. We conjecture that the radiative energy thus adsorbed in the crystal is diffused towards the underlying film as athermal phonons, and that this is a low efficiency process due to the mechanical mismatch between the soft molecular crystals, the amorphous oxide, and the metallic grains. Work on superconducting qubits suggests about an order of magnitude smaller effective blackbody radiation temperature \cite{serniak2018hot} impinging on devices protected by standard infrared filtering. While we expect that the shielding mechanism due to the hydrogen crystals remains effective, the amount to which this will correspond to an improvement in the quality factor of devices will depend on the weight of quasiparticle dissipation in the loss budget. Finally, we would like to draw a parallel to Ref.~\cite{Lucas2023}, in which a similarly high quality dielectric is used to improve the coherence of a quantum circuit | albeit via a different mechanism | while not introducing measurable losses.

% relevance and outlook
These results have both fundamental and practical relevance. Firstly, they indicate a possible nontrivial light-matter interaction, which may be exploited to protect superconducting devices from terahertz radiation by coating them with a thin, soft, low loss dielectric. Secondly, they provide a step towards realizing hybrid quantum architectures with impurities embedded into magnetically inert and low dielectric loss matrices grown onto high impedance superconducting resonators~\cite{Albertinale2021, Wang2023}. 

\subsection*{Data availability}
All relevant data are available from the authors upon reasonable request.

\subsection*{Competing interest}
The authors declare no competing interest.

\section*{Acknowledgements}
We acknowledge A.~Monfardini, M.~Calvo and F.~Lévy-Bertrand for providing the resonator design and the sample holders, W.~Wulfhekel for fruitful discussions regarding surface effects, T. Reisinger and S. Günzler for critical proof-reading of the manuscript draft, and L.~Radtke, S.~Diewald and A.~Lukashenko for technical support. Facilities use is supported by the KIT Nanostructure Service Laboratory (NSL). F.V. and I.M.P. acknowledge support from the German Ministry of Education and Research (BMBF) within the project GEQCOS (FKZ: 13N15683). A.N.K, A.A. and J.S. acknowledge support by the Austrian Science Fund (FWF) project P34314 (Spins in Quantum Solids), as well as the European Union’s Horizon 2020 research and innovation programme (FET- OPEN project FATMOLS, Grant No. 862893). L.B.-I. and C.M. acknowledge support from UEFISCDI Romania through the project ERANET-QUANTERA-QuCos 120/16.09.2019 and from The Romanian Ministry of Research through NUCLEU program, project PN23 24 01 04.

\setcounter{figure}{0}
\renewcommand{\thefigure}{S\arabic{figure}}
\renewcommand{\theHfigure}{A\arabic{figure}}

\section*{Appendices} \label{sec_suppl}
\appendix

 \begin{figure}[t!]
  \def\svgwidth{1\columnwidth} 
    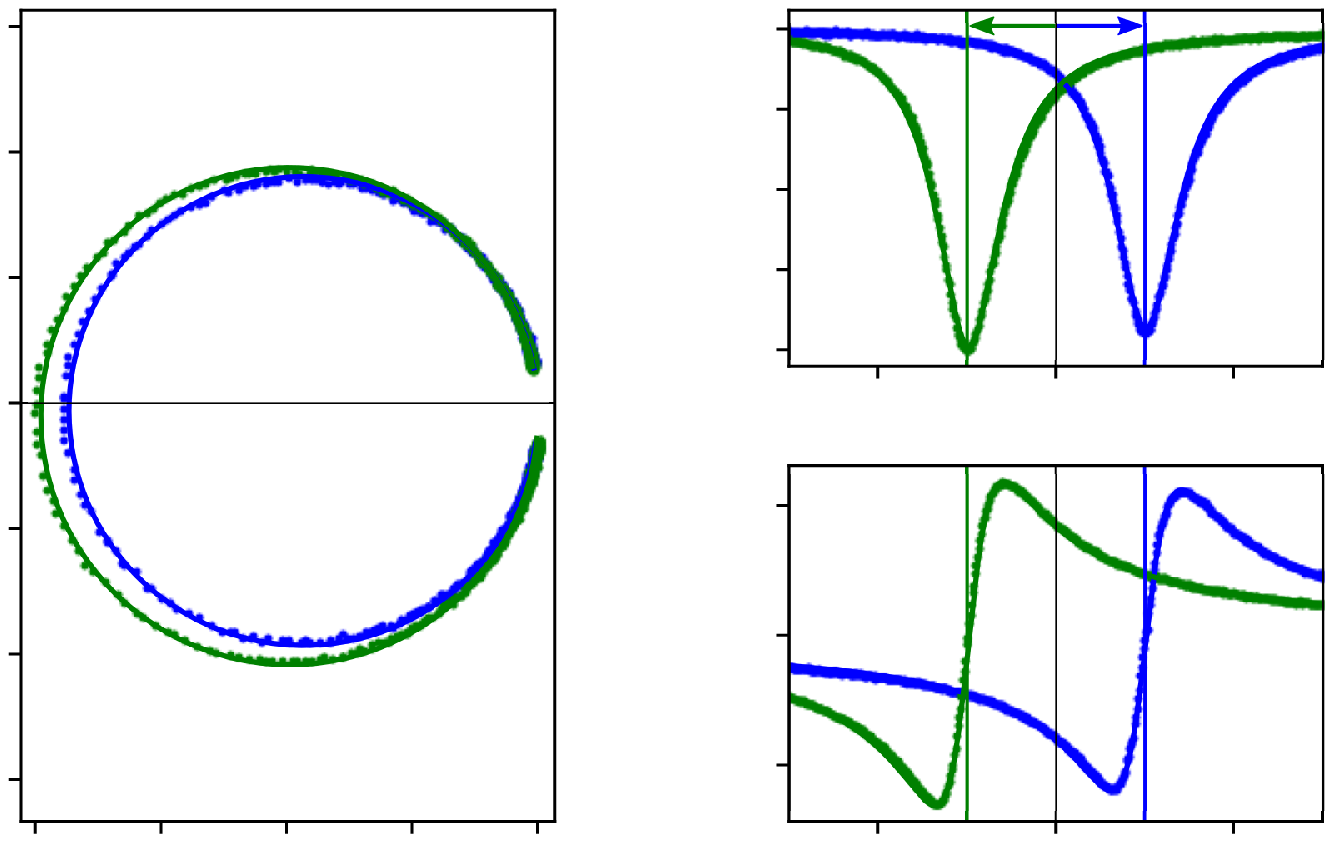
    \caption{Scattering data (markers) and fit (solid lines) to the circle fit model used to extract microwave parameters, shown for a resonator in sample S with the 1 cm aperture before (blue) and after (green) deposition of a thick (saturating) crystal. Note that the transmission data is normalized to the sample holder response, and the frequency response in the right hand side panels is horizontally offset by a linewidth $\kappa = f_0/Q \sim 25$~kHz about the resonance for clarity, where $1/Q = 1/Q_i + 1/Q_c$ and $Q_c=1.66\times 10^5$.} \label{fig_circle}
\end{figure}

\begin{figure*}[t!]
  \def\svgwidth{1\textwidth} 
    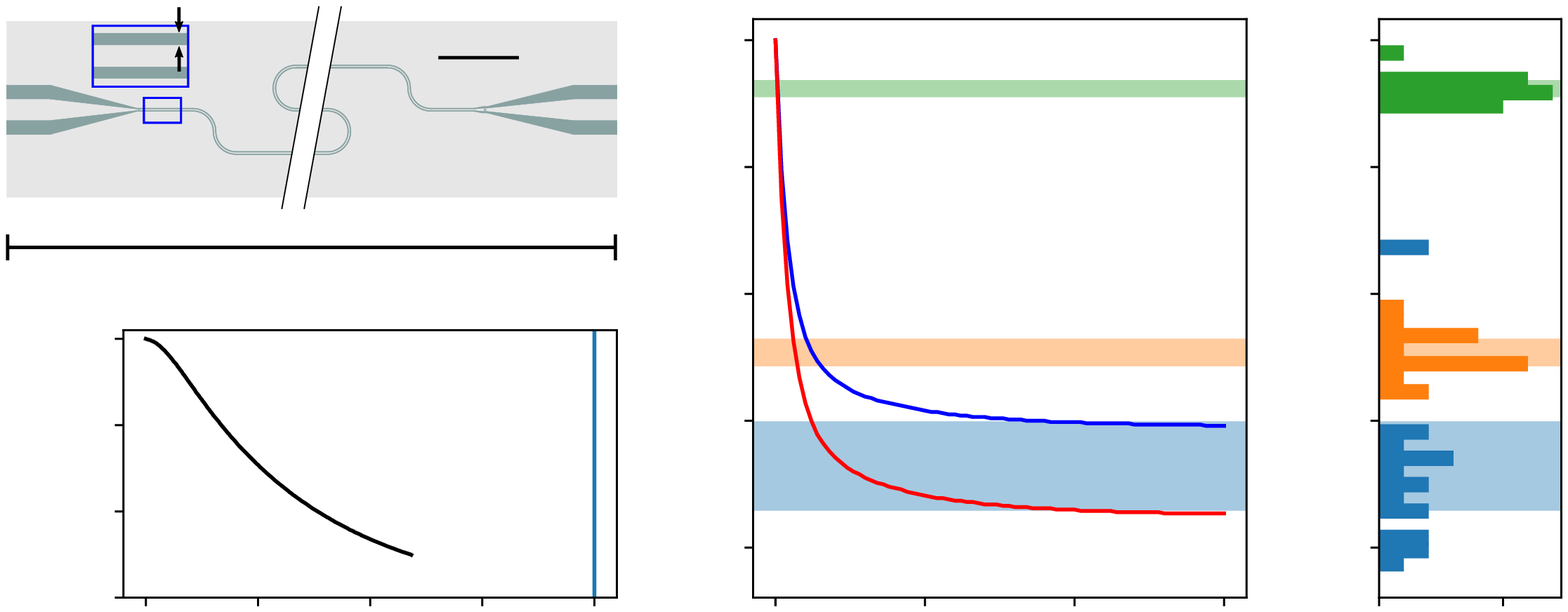
    \caption{\textbf{Crystal thickness calibration}. \textbf{a}, drawing of the niobium short-ended $\lambda/2$ ``calibrator'' resonator. The pitch to ground of the stripline ($8\;\upmu$m) is comparable to that between fingers of the interdigitated capacitors of samples S and L ($12\;\upmu$m; cf.~Fig.~\ref{fig_loading}a). \textbf{b}, real-time evolution of the resonant frequency of the Nb stripline during hydrogen deposition. \textbf{c}, simulated (solid lines) and \textbf{d}, measured (histograms) resonant frequency shift for the grAl resonators in sample S. Measured values refer to the setup with a 1 cm hole (cf.~Fig.~\ref{fig_qi}); the shaded regions spanning both panels cover one standard deviation about the mean of measured values. By using the ``thick'' parameter set as a calibration for a crystal fully saturating the dielectric shift we obtain a permittivity for the para-hydrogen crystal of $1.38\pm0.04$, and we calibrate two other parameters sets, ``thin'' and ``very thin'', to result in a crystal thickness of order of magnitude $10$ and $1\;\upmu$m, respectively.} \label{fig_calibration}
\end{figure*}

\begin{figure}[t!]
  \def\svgwidth{1\columnwidth} 
    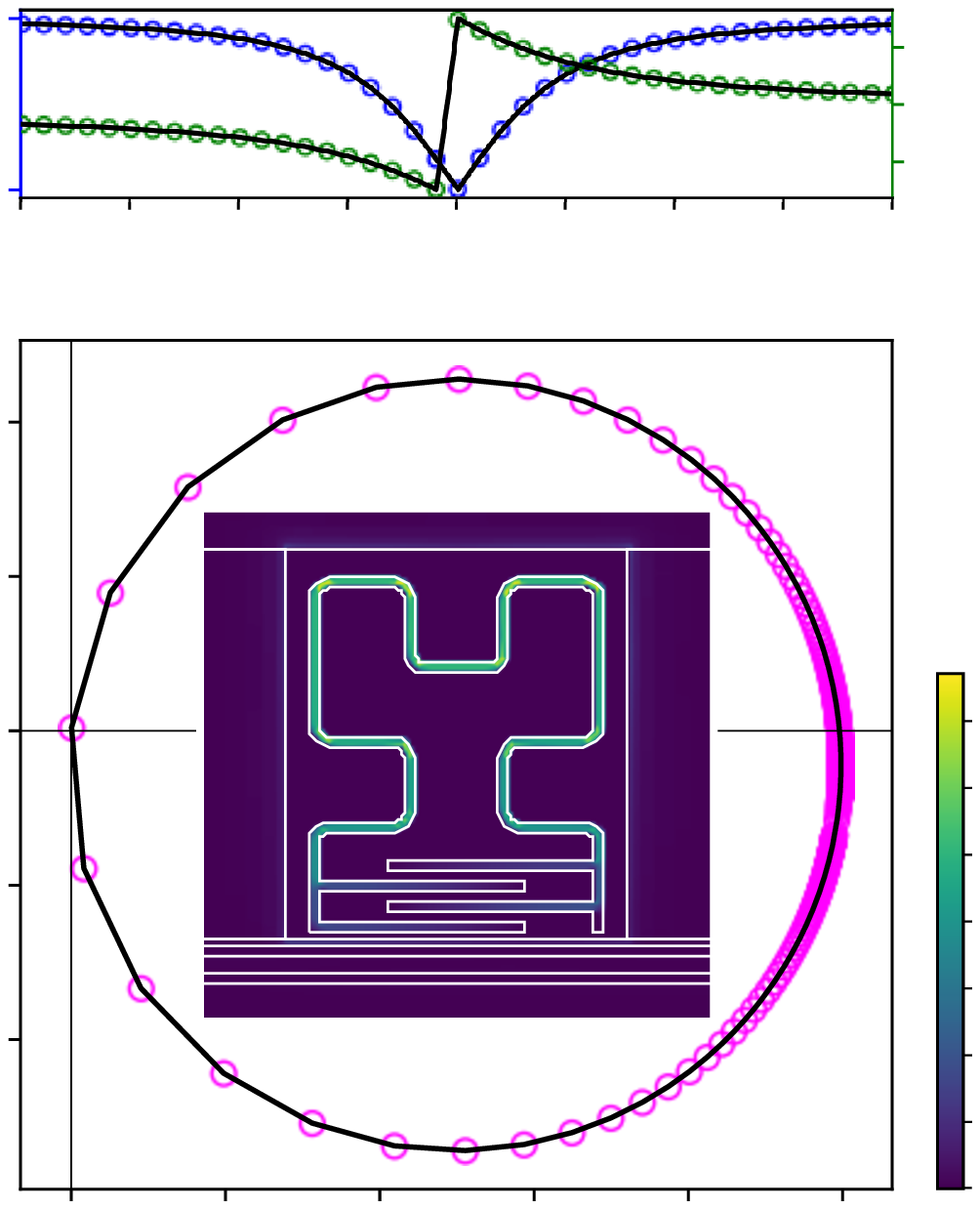
    \caption{Results of Sonnet simulation of a single resonator from sample S. The microwave response is shown in amplitude and phase (top panel) and as a circle in the complex plane (bottom panel). Open circles show the simulation data. The solid black line is the fit to the data. The inset shows the computed amplitude of surface current density on resonance, with overlayed geometry in white lines.} \label{fig_sonnet}
\end{figure}

\begin{figure*}[t]
  \def\svgwidth{0.99\textwidth}     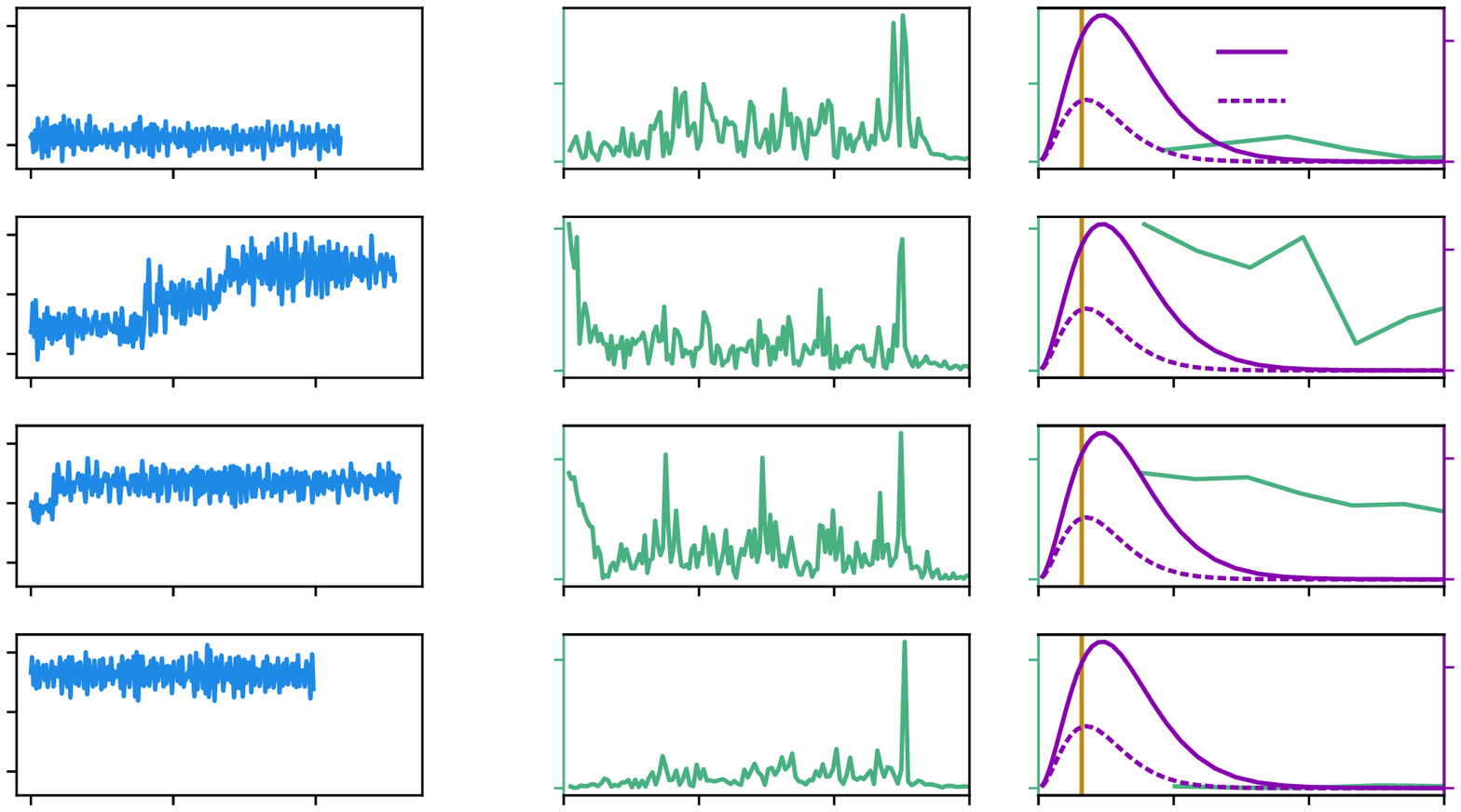
  \caption{Examples of time traces and relative spectra for increasing amounts of hydrogen molecules (top to bottom). Left: time series of the supercell dipole moment evolution. Center: spectral content of the dipole moment dynamics. Right: comparison between the (zoomed in) spectral content from the central panel (green), the blackbody radiation from the gas tube estimated to be between 3 and 4 K (purple), and the superconducting spectral gap (gold). Visible fluctuations were observed only for $N=5$ and $6$ for purely statistical reasons | the number of molecules does not influence the supercell dynamics.}\label{fig_series_and_spectra}
 \end{figure*} 

\section{Resonance circle fit} \label{sec_circle_fit}
We show an example of a resonator response before and after crystal deposition in Fig.~\ref{fig_circle}. We highlight two figures of merit: the resonant frequency, decreased by $\sim 30$ MHz due to the crystal saturating electric field lines above the grAl film, and $Q_i$ increased by roughly 10\%. 

\section{Photon number calibration} 
\label{sec_phot_num}

We perform power calibration to obtain the average number of circulating photons in the resonators using $\bar{n} = 2 Q^2 P_\mathrm{cold} / ( Q_c \hbar \omega _0^2)$, where $\omega_0$ is the resonant frequency in radians per second, $Q$ and $Q_c$ are the total and coupling quality factors, and $P _\mathrm{cold}$ is the microwave power at the input port. The latter is computed by estimating the total attenuation of the line. The total attenuation, measured at room temperature from the cryostat input to the on-sample port, is about $65$~dB in the $2-6$~GHz range, indicating $15$~dB of attenuation due to wiring added to the nominal $50$~dB (cf.~Fig.~\ref{fig_setup}a).

\section{Crystal thickness calibration} \label{sec_thickness}

In order to calibrate the thickness of the deposited crystals, we firstly perform measurements on a $\lambda/2$ short-ended resonator made of niobium (cf.~Fig.~\ref{fig_calibration}a). The critical temperature of Nb, $T_c \sim 9$~K, allows one to measure the resonator during the hydrogen deposition. This is in contrast to grAl, which has a critical temperature of circa $2$~K and transitions to the normal state during hydrogen deposition (at which point the temperature of the chip, due to the gas flow, reaches $3$~K). 

We regulate the needle valve v$_1$ (cf.~Fig.~\ref{fig_setup}a) until the incoming pressure from the room temperature reservoir reads $11$~mBar on the digital barometer and measure successive traces around the resonant frequency. In Fig.~\ref{fig_calibration}b we plot the measured decrease in resonant frequency due to the added dielectric. After a few minutes, the curve starts to saturate as the deposited crystal outgrows the relatively localized electric field lines of the resonator. Deposition at $11$~mBar for $20$~minutes saturates the frequency shift of the resonator at $30$~MHz. Assuming that electric field lines out of the sapphire plane are fully confined in the hydrogen crystal, we estimate $\epsilon_{\mathrm{p-H_2}} = 1.23$. This is in acceptable agreement with the value $\epsilon_{\mathrm{p-H_2}} = 1.3$ reported by \cite{younglove} at $14$~K and kHz frequency. While we do observe an increase in the total quality factor for the Nb resonator after crystal deposition, which indicates an improvement of $Q_i$, we can not make a quantitative statement. The reason is that we only measured one such device and the Nb resonator is coupled in a band-pass configuration in which disentangling $Q_i$ and $Q_c$ is prone to artifacts. In order to be quantitative, statistics should be acquired on tens of resonators in reflection or notch configurations.

We turn our attention to the grAl resonators. The only assumption here is that the same growth parameter set which resulted in a crystal that saturated the frequency shift for the Nb stripline will also saturate the grAl resonators shift. This assumption is justified by the similar capacitor pitches ($8$ and $12\;\upmu$m, respectively) and correspondingly similar out of plane electric field extension for the two geometries. As shown in Fig.~\ref{fig_calibration}c and d, we measured on average $33\pm7$~MHz of frequency shift after a ``saturating'' deposition on grAl, corresponding to crystals with $\epsilon_{\mathrm{p-H_2}} = 1.38\pm0.04$, again comparable to literature, and thickness above $100\;\upmu$m. Equipped with this value, we calibrate two other sets of growth parameters, indicating crystals on the order of $10$ and $1\;\upmu$m thick. The surface impedance of this new, magnetically inert interface with vacuum is thus $Z=\sqrt{\upmu_0/(1.38\epsilon_0)}=320 \; \Omega$, giving a total power reflection $\Gamma^2 = ((Z-Z_0)/(Z+Z_0))^2 \lessapprox 1\% $, allowing us to rule out introduced reflections of the black body radiation at the hydrogen-vacuum interface as the source of the $Q_i$ improvement.

\section{Sonnet simulations} \label{sec_sonnet_simu}

We use the Sonnet software in order to simulate the effect of adding the hydrogen molecules. 
Both superconducting films | aluminum and high kinetic inductance granular aluminum | are modelled as lossless metal with an added sheet inductance, calculated from room temperature DC transport values according to Mattis-Bardeen theory \cite{Rotzinger_2016}. The added para-hydrogen crystal is modelled as a uniform dielectric of variable thickness and permittivity. Metalizations are discretized in $4\times4 \;\upmu$m$^2$ cells. The total cell, comprising the resonator, feedline, and a part of the ground plane, measures $600 \times 600 \; \upmu$m$^2$. Results of a single simulation are shown in Fig.~\ref{fig_sonnet}, for a resonator in sample S with a $1\;\upmu$m thick hydrogen layer having permittivity $\epsilon = 1.22$. The resulting data can be fitted to extract the resonant frequency. The fitted coupling quality factor $Q_c = 2.1 \times 10^5$ is in agreement with the measured data, whereas the internal quality factor is essentially infinite (notice the amplitude reaching zero) since all materials are modeled as lossless.

The resulting resonant frequency $f_0=2.283$~GHz is lower than all measured resonances for that sample | roughly in the $2.4-3.2$~GHz range | possibly due to the rough estimation of the kinetic inductance. To compute the current density we set a drive voltage of $4\;\upmu$V at the input port, which is the root-mean-square equivalent to the on-chip power fed to resonators in Fig.~\ref{fig_qi}, estimated to be $-95$~dBm. Note the current distribution indicative of a lumped element structure, with the capacitor fully charged and uniform flow in the inductive meander, and the current magnitude, well below the critical value (estimated at $\sim 1$~mA/$\upmu$m$^2$ for highly oxidized grAl \cite{Maleeva2018}).

\section{Details on DFT simulations} \label{sec_dft}

\subsection{Methods and model}

We have calculated the properties of the systems (H$_2$ molecules on Al$_2$O$_3$ surface) using the {\sc Siesta} code \cite{ordejon_self,siesta_code} that uses norm-conserving pseudopotentials in their fully non-local form and expands the wavefunctions of valence electrons by flexible linear combinations of atomic orbitals (LCAO). 
A double-zeta polarized basis set has been employed for all atoms. 
The alumina supercell is a periodically repeated slab, consisting of 72 aluminum atoms and 108 oxygen atoms according to the supercell of $3\times 3\times 2$ of the unit-cell structure provided by the Materials Project \cite{materialsproj}.
The system that includes the hydrogen molecules has a $z$-axis size of 30 \AA; this allows a vacuum level of $\approx$ 20 \AA\ between the cells. Such a large cell eliminates the interactions between periodically repeated images of the adsorbed molecules and avoids the artificial influence of the electric charge from one cell to another.
We used a $2\times 2$ Monkhorst-Pack grid \cite{monkhorst_pack_grid} for the integrals in the Brillouin zone for the transversal direction while the periodicity along the $z$-axis was modeled with a single $k$-point.  
The vdW-DF-cx exchange-correlation functional of Berland and Hyldgaard (BH) \cite{berland_hyldgaard_functional} was used in order to get information on Van der Waals components of the forces.
 We ran our computation on the INCDTIM Datacenter on an IBM System with 14 Blades Quad Core Dual Processor (each operating on 16 GB RAM at 2.5 GHz).

\subsection{Adsorption geometry}

We run separate simulations, with 1, 2, 3, 4, 5, 6 and 8 H$_2$ hydrogen  molecules on top of aluminum oxide. Molecules are added randomly in the vacuum layer atop the supercell, at an average distance of $2$~\r{A}. The ensuing relaxation process is iterated via DFT calculations until the gradients fall below the $10$~meV/\r{A} threshold. The DFT calculations indicate two distinct adsorption geometries, with the interatomic axis either almost perpendicular ($\sim 80 ^\circ $ tilting angle) or almost parallel ($\sim 15 ^\circ $ tilting angle to the supercell surface), which we refer to as \hperp~and \hpar, respectively. The single molecule ($N=1$) relaxes to an average distance of $1.8$~\r{A} from the surface, indicating the presence of electrostatic interactions, whereas the $2$ to $8$ cases relax to  $ \sim 2.3$~\r{A}, suggesting a dominant van der Waals interaction. Most probably the two distinct adsorption geometries are due to the surface structure of alumina: \hpar~bridges two oxygen atoms over an aluminum atom, while the \hperp~falls in the triangle made of three oxygen atoms (cf.~Fig.~\ref{fig_ir}a | top view).

While the parallel orientation \hpar~is energetically more stable compared to \hperp, the absolute differences in adsorption energies are small and can be overcome by terahertz excitation, changes in the local environment from other adsorbates, etc. Consequently, switching between the two orientations is possible during the experiment. Moreover, this switch is reflected in the molecule-surface charge transfer, which ultimately determines discrete-level fluctuations of the electrical dipole at the surface. Since the terahertz spectrum is the Fourier transform of the dipolar momentum and the latter exhibits step-functions, we expect significant terahertz absorption at relatively low frequencies, in the range of hundreds of GHz, comparable to the superconducting spectroscopic gap.

\subsection{Charge transfer}

By plotting the difference between the charge of single atoms and the actual surface charge of the supercell, a significant spatial modulation of the oxidation numbers can be appreciated (cf.~Fig.~\ref{fig_ir}a | isosurface on the left). The adsorption selectivity (i.e between \hpar~and \hperp) is correlated to the corresponding charge transfer defined by\begin{align} \label{eq_charge_trans}
    \Delta \rho (\bm{r}) &= \rho _\text{Adsorbate/Substrate}  \nonumber \\  &- [ \rho _\text{adsorbate }(\bm{r})  + \rho _\text{substrate} (\bm{r}) ].
\end{align}
The three right-hand-side terms are: the charge densities of the complete and fully relaxed system, the adsorbate only, and the clean surface only, respectively.
As shown by the isocharge clouds in Fig.~\ref{fig_ir}a (positive in olive-green and negative in blue), the different adsorption geometries result in either a molecule $\to$ surface (\hpar) or surface $\to$ molecule (\hperp) charge transfer. The charge imbalance is of the order of one electron every thousands cubic \r{A}, i.e. a per mille perturbation of the hydrogen molecule charge.

\subsection{Molecular dynamics}

To calculate the dipolar momentum for each system, we perform a molecular dynamics simulation. The equations of motion are solved with a Verlet algorithm with a $1.5$~fs step, for roughly $3000$ steps. The values of the dipolar momentum/supercell for the models with $N=4, 5, 6, 8$ molecules on surface as well as the terahertz spectra calculated as Fourier transforms of the dipole are given in Fig.~\ref{fig_series_and_spectra}. The system is insofar assumed to be microcanonical; in reality, perturbations of the system can result in an activity in these degrees of freedom, resulting in the charged dipoles fluctuating over time.

%\bibliography{main_refe.bib}

\end{document}